# An Empirical Comparative Study of Checklist-based and Ad Hoc Code Reading Techniques in a Distributed Groupware Environment

**Olalekan S. Akinola**
Department of Computer Science,
University of Ibadan, Nigeria.
Solom202@yahoo.co.uk

**Adenike O. Osofisan**
Department of Computer Science,
University of Ibadan, Nigeria
mamoshof@yahoo.co.uk

## Abstract

Software inspection is a necessary and important tool for software quality assurance. Since it was introduced by Fagan at IBM in 1976, arguments exist as to which method should be adopted to carry out the exercise, whether it should be paper-based or tool-based, and what reading technique should be used on the inspection document. Extensive works have been done to determine the effectiveness of reviewers in paper-based environment when using ad hoc and checklist reading techniques. In this work, we take the software inspection research further by examining whether there is going to be any significant difference in defect detection effectiveness of reviewers when they use either ad hoc or checklist reading techniques in a distributed groupware environment. Twenty final year undergraduate students of computer science, divided into ad hoc and checklist reviewers groups of ten members each were employed to inspect a medium-sized java code synchronously on groupware deployed on the Internet. The data obtained were subjected to tests of hypotheses using independent t-test and correlation coefficients. Results from the study indicate that there are no significant differences in the defect detection effectiveness, effort in terms of time taken in minutes and false positives reported by the reviewers using either ad hoc or checklist based reading techniques in the distributed groupware environment studied.

## I. INTRODUCTION

A software could be judged to be of high or low quality depending on who is analyzing it. Thus, quality software can be said to be a *"software that satisfies the needs of the users and the programmers involved in it"*, [28]. Pfleeger highlighted four major criteria for judging the quality of a software:

(i) It does what the user expects it to do;
(ii) Its interaction with the computer resources is satisfactory;
(iii) The user finds it easy to learn and to use; and
(iv) The developers find it convenient in terms of design, coding, testing and maintenance.

In order to achieve the above criteria, software inspection was introduced. Software inspection has become widely used [36] since it was first introduced by Fagan [25] at IBM. This is due to its potential benefits for software development, the increased demand for quality certification in software, (for example, ISO 9000 compliance requirements), and the adoption of the Capability Maturity Model as a development methodology [27].

Software inspection is a necessary and important tool for software quality assurance. It involves strict and close examinations carried out on development products to detect defects, violations of development standards and other problems [18]. The development products could be *specifications, source code, contracts, test plans and test cases* [33, 4, 8].

Traditionally, the software inspection artifact (requirements, designs, or codes) is normally presented on papers for the inspectors / reviewers. The advent of Collaborative Software Development (CSD) provides opportunities for software developers in geographically dispersed locations to communicate, and further build and share common knowledge repositories [13]. Through CSD, distributed collaborative software inspection methodologies emerge in which group of reviewers in different geographical locations may log on synchronously or asynchronously online to inspect an inspection artifact.

It has been hypothesized that in order to gain credibility and validity, software inspection experiments have to be conducted in different environments, using different people, languages, cultures, documents, and so on [10, 12]. That is, they must be redone in some other environments. The motivation for this work therefore stems from this hypothesis.

Specifically, the following are the target goals of this research work, to determine if there is any significant difference in the effectiveness of reviewers using ad hoc code reading technique and those using Checklist reading technique in a distributed tool-based environment.

Twenty final year students of Computer Science were employed to carry out inspection task on a medium-sized code in a distributed, collaborative environment. The students were grouped into two groups. One group used the Ad hoc code reading technique while the second group used the checklist-based code reading technique (CBR). Briefly, results obtained show that there is no significant difference in the effectiveness of reviewers using Ad hoc and Checklist reading technique in the distributed environment.

In the rest of this paper, we focus on review of related work in section 2. In section 3, we stated the experimental planning and instruments used as well as the subjects and hypotheses set up for the experiment. Threats to internal and external validities are also treated in this section. In section 4, we discuss the results and statistical tests carried out on the data in the experiment. Section 5 is about the discussion of our results while section 6 states the conclusion and recommendations.

## II. Related Works: Ad hoc versus Checklist-based Reading Techniques

Software inspection is as old as programming itself. It was introduced in the 1970s at IBM, which pioneered its early adoption and later evolution [25]. It is a way of detecting faults in a software documents – requirements, design or code. Recent empirical studies demonstrate that defect detection is more an individual than a group activity as assumed by many inspection methods and refinements [20, 29, 22]. Inspection results depend on inspection participants themselves and their strategies for understanding the inspected artifacts [19].

A defect detection or reading (as it is popularly called) technique is defined as the series of steps or procedures whose purpose is to guide an inspector in acquiring a deep understanding of the inspected software product [19]. The comprehension of inspected software products is a prerequisite for detecting subtle and / or complex defects, those often causing the most problems if detected in later life cycle phases.

According to Porter *et al*, [1], defect detection techniques range in prescription from intuitive, nonsystematic procedures such as ad hoc or checklist techniques, to explicit and highly systematic procedures such as scenarios or correctness proofs. A reviewer's individual responsibility may be general, to identify as many defects as possible, or specific, to focus on a limited set of issues such as ensuring appropriate use of hardware interfaces, identifying un-testable requirements, or checking conformity to coding standards.

Individual responsibilities may or may not be coordinated among the review team members. When they are not coordinated, all reviewers have identical responsibilities. In contrast, each reviewer in a coordinated team has different responsibilities.

The most frequently used detection methods are ad hoc and checklist. Ad-hoc reading, by nature, offers very little reading support at all since a software product is simply given to inspectors without any direction or guidelines on how to proceed through it and what to look for. However, ad-hoc does not mean that inspection participants do not scrutinize the inspected product systematically. The word `ad-hoc' only refers to the fact that no technical support is given to them for the problem of how to detect defects in a software artifact. In this case, defect detection fully depends on the skill, the knowledge, and the experience of an inspector. Training sessions in program comprehension before the take off of inspection may help subjects develop some of these capabilities to alleviate the lack of reading support [19].

Checklists offer stronger, boilerplate support in the form of questions inspectors are to answer while reading the document. These questions concern quality aspects of the document. Checklists are advocated in many inspection works. For example, Fagan [24,25], Dunsmore [2], Sabaliauskaite [12], Humphrey [35] and Gilb and Grahams' manuscript [32] to mention a few.

Although reading support in the form of a list of questions is better than none (such as ad-hoc), checklist-based reading has several weaknesses [19]. First, the questions are often general and not sufficiently tailored to a particular development environment. A prominent example is the following question: "Is the inspected artifact correct?" Although this checklist question provides a general framework for an inspector on what to check, it does not tell him or her in a precise manner how to ensure this quality attribute. In this way, the checklist provides little support for an inspector to understand the inspected artifact. But this can be vital to detect major application logic defects. Second, how to use a checklist are often missing, that is, it is often unclear when and based on what information an inspector is to answer a particular checklist question. In fact, several strategies are actually feasible to address all the questions in a checklist. The following approach characterizes the one end of the spectrum: The inspector takes a single question, goes through the whole artifact, answers the question, and takes the next question. The other end is defined by the following procedure: The inspector reads the document. Afterwards he or she answers the questions of the checklist. It is quite unclear which approach inspectors follow when using a checklist and how they achieved their results in terms of defects detected. The final weakness of a checklist is the fact that checklist questions are often limited to the detection of defects that belong to particular defect types. Since the defect types are based on past defect information, inspectors may not focus on defect types not previously detected and, therefore, may miss whole classes of defects.

To address some of the presented difficulties, one can develop a checklist according to the following principles [19]:

- The length of a checklist should not exceed one page.
- The checklist question should be phrased as precise as possible.
- The checklist should be structured so that the quality attribute is clear to the inspector and the question give hints on how to assure the quality attribute. And additionally,
- The checklist should not be longer than a page approximately 25 items [12, 32].

In practice, reviewers often use Ad Hoc or Checklist detection techniques to discharge identical, general responsibilities. Some authors, especially Parnas and Weiss [9] have argued that inspections would be more effective if each reviewer used a different set of systematic detection techniques to discharge different specific responsibilities.

Computer and / or Internet support for software inspections has been suggested as a way of removing the bottlenecks in the traditional software inspection process. The web approach makes software inspection much more elastic, in the form of asynchronicity and geographical dispersal [14].

The effectiveness of manual inspections is dependent upon satisfying many conditions such as adequate preparations, readiness of the work product for review, high quality moderation, and cooperative interpersonal relationships. The effectiveness of tool-based inspection is less dependent upon these human factors [29, 26]. Stein *et al.*, [31] is of the view that distributed, asynchronous software inspections can be a practicable method. Johnson [17] however opined that thoughtless computerization of the manual inspection process may in fact increase the cost of inspections.

To the best of our knowledge, many of the works in this line of research from the literatures either report experiences in terms of lessons learned with using the tools, for instance, Harjumaa [14] and Mashayekhi [34], or compare the effectiveness of tools with paper-based inspections, for instance, Macdonald and Miller [23]. In the case of ICICLE [7], the only published evaluation comes in the form of lessons learned. In the case of Scrutiny, in addition to lessons learned [16], the authors also claim that tool-based inspection is as effective as paper-based, but there is no quantifiable evidence to support this claim [15].

In this paper, we examine the feasibility of tool support for software code inspection as well as determining if there is any significant difference in the effectiveness of reviewers using ad hoc and checklist reading techniques in a distributed environment.

## III. Experimental Planning and Design

### A. Subjects:

Twenty 20 final year students of Computer Science were employed in the study. Ten (10) of the student-reviewers used Ad Hoc reading technique, without providing any aid for them in the inspection. The other ten used Checklist based reading technique. The tool provided them with some checklist as aid for the inspection.

### B. Experimental Instrumentation and Set up

Inspro, a web-based distributed, collaborative code inspection tool was designed and developed as a code inspection groupware used in the experiment. The experiment was actually run in form of synchronous, distributed collaborative inspection with the computers on the Internet. One computer was configured as a server having WAMP server installed on it, while the other computers served as clients to the server.

The tool was developed by the authors, using Hypertext Preprocessor (PHP) web programming language and deployed on Apache Wamp Server. The student reviewers were orientated on the use of the Inspro web-based tool as well as the code artifact before the real experiment was conducted on the second day. The tool has the following features:

(i) The user interface is divided into three sections. A section displays the code artifact to be worked upon by the reviewers along with their line numbers, while another section displays the text box in which the reviewers keyed in the bugs found in the artifact. The third section below is optionally displayed to give the checklist to be used by the reviewers if they are in checklist group and does not display anything for the ad hoc inspection group.

(ii) Immediately a reviewer log on to the server, the tool starts counting the time used for the inspection and finally records the time stopped when submit button is clicked.

(iii) The date the inspection is done as well as the prompt for the name of the reviewer is automatically displayed when the tool is activated.

(iv) The tool actually writes the output of the inspection exercise on a file which will be automatically opened when submit button is clicked. This file will then be finally opened and printed for further analysis by the chief coordinator of the inspection.

Fig. 1 displays the user interface of the Inspro tool.

### C. Experimental Artifact

The artifact used for this experiment was a 156 lines java code which accepts data into two 2-dimensional arrays. This small-sized cod was used because the students involved in the experiment had their first experience in code inspection with this experiment; even though they were given some formal trainings on code inspection prior the exercise. The experiment was conducted as a practical class in a Software Engineering course in 400 level (CSC 433) in the Department of Computer Science, University of Ibadan, Nigeria. The arrays were used as matrices. Major operations on matrices were implemented in the program such as sum, difference, product, determinant and transpose of matrices. All conditions for these operations were tested in the program. The code was developed and tested okay by the researcher before it was finally seeded with 18 errors: 12 logical and 6 syntax/semantic. The program accepts data into the two arrays, then performs all the operations on them and reports the output results of computation if there were no errors. If there were errors in form of operational condition not being fulfilled for any of the operations, the program reports appropriate error log for that operation.

### D. Experimental Variables and Hypothesis

The experiment manipulated 2 independent variables: The number of reviewers per team (1, 2, 3, or 4 reviewers, excluding the code author) and the review method – ad hoc and checklist. Three dependent variables were measured for the independent variables: the average number of defects detected by the reviewers, that is, defect detection effectiveness (DE), the average time spent for the inspection (T) in minutes and the average number of false positives reported by the reviewers (FP). The defect detection effectiveness (DE) is the number of true defects detected by reviewers out of the total number of seeded defects in a code inspection artifact. The time measures the total time (in minutes) spent by a reviewer on inspecting an inspection artifact. Inspection Time is also a measure of the effort (E) used by the reviewers in a software inspection exercise. False positives (FP) are the perceived defects a reviewer reported but are actually not true defects.

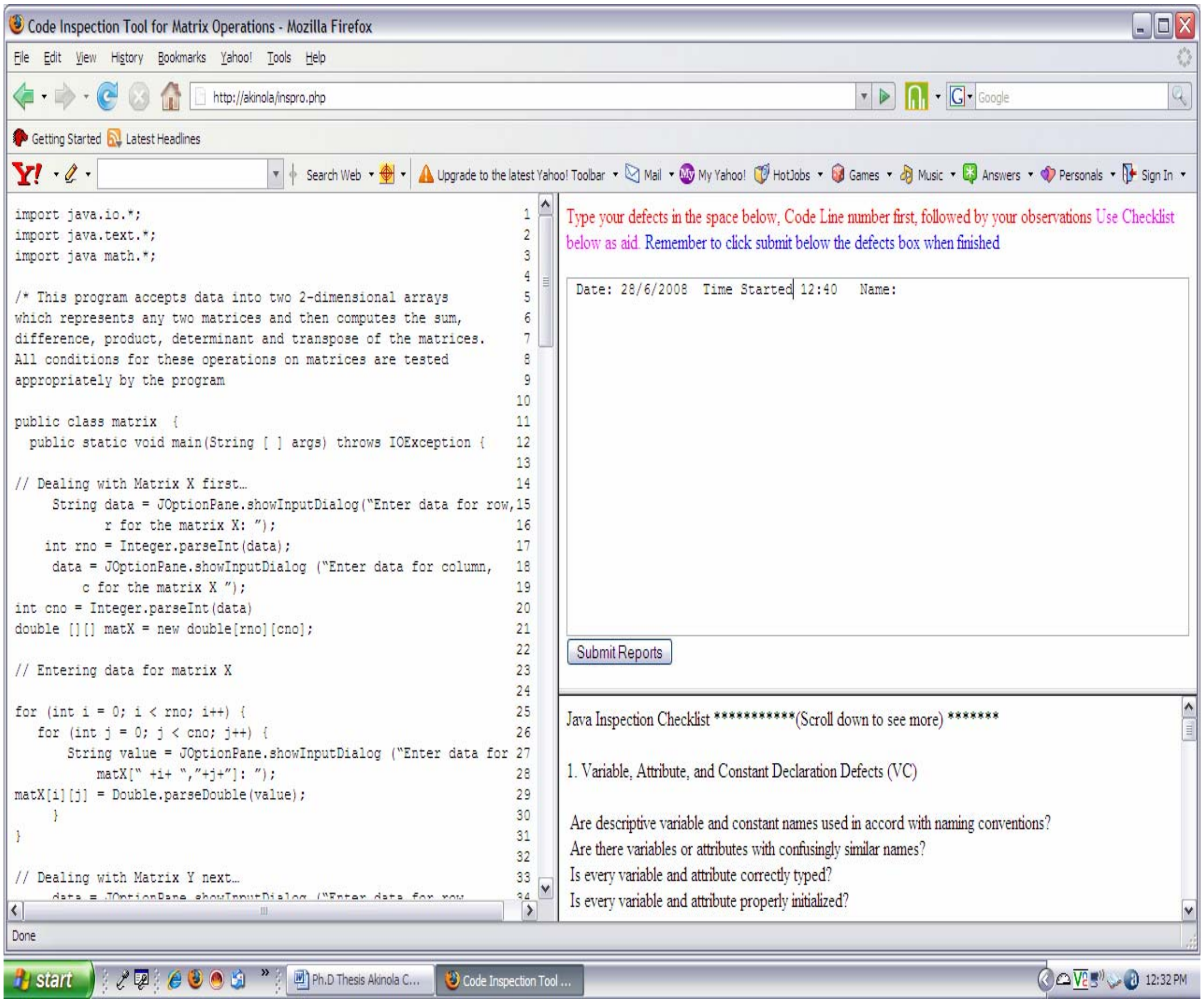

**Fig.1 Inspro User Interface**

Three hypotheses were stated for this experiment as follows.

$H_o1$: There is no significant difference between the effectiveness of reviewers using Ad hoc and Checklist reading techniques in distributed code inspection.

$H_o2$: There is no significant difference between the effort taken by reviewers using Ad hoc and Checklist techniques in distributed code inspection.

$H_o3$: There is no significant difference between the false positives reported by reviewers using Ad hoc and Checklist techniques in distributed code inspection.

### E. Threats to validity

The question of validity draws attention to how far a measure really measures the concept that it purports to measure (Alan and Duncan, 1997). Therefore, in this experiment, we considered two important threats that may affect the validity of the research in the domain of software inspection.

### F. Threats to Internal Validity

Threats to internal validity are influences that can affect the dependent variables without the researcher's knowledge. We considered three such influences: (1) selection effects, (2) maturation effects, and (3) instrumentation effects.

Selection effects are due to natural variation in human performance [1]. For example, if one-person inspections are done only by highly experienced people, then their greater than average skill can be mistaken for a difference in the effectiveness of the treatments. We limited this effect by randomly assigning team members for each inspection. This way, individual differences were spread across all treatments.

Maturation effects result from the participants' skills improving with experience. Randomly assigning the reviewers and doing the review within the same period of time checked these effects.

Instrumentation effects are caused by the artifacts to be inspected, by differences in the data collection forms, or by other experimental materials. In this study, this was negligible or did not take place at all since all the groups inspected the artifacts within the same period of time on the same web based tool.

### 3.4.2 Threats to External Validity

Threats to external validity are conditions that can limit our ability to generalize the results of experiments to industrial practice [1]. We considered three sources of such threats: (1) experimental scale, (2) subject generalizability, and (3) subject and artifact representativeness.

Experimental scale is a threat when the experimental setting or the materials are not representative of industrial practice. This has a great impact on the experiment as the material used (matrix code) was not a true representative of what obtains in industrial setting. The code document used was invented by the researchers. A threat to subject generalizability may exist when the subject population is not drawn from the industrial population. We tried to minimize this threat by incorporating 20 final year students of Computer Science who have just concluded a 6 months industrial training in the second semester of their 300 level. The students selected were those who actually did their industrial trainings in software development houses.

Threats regarding subject and artifact representativeness arise when the subject and artifact population is not representative of the industrial population. The explanations given earlier also account for this threat.

## 4. Results

Table 1 shows the raw results obtained from the experiment. The “defect (%)” gives the percentage true defects reported by the reviewers in each group (Ad hoc and Checklist) from the 18 seeded errors in the code artifact, the “effort” gives the time taken in minutes for reviewers to inspect the online inspection documents while the “No of FP” is the total number of false positives reported by the reviewers in the experiment.

**Table 1: Raw Results Obtained from Collaborative Code Inspection**

| | **Ad hoc Inspection** | | | | **Checklist Inspection** | | |
|---|---|---|---|---|---|---|---|
| s/n | Defect (%) | Effort (Mins) | No of FP | | Defect (%) | Effort (Mins) | No of FP |
| 1 | 27.78 | 31 | 1 | | 44.44 | 80 | 5 |
| 2 | 44.44 | 72 | 5 | | 33.33 | 60 | 5 |
| 3 | 50.00 | 57 | 4 | | 27.78 | 35 | 4 |
| 4 | 50.00 | 50 | 2 | | 22.22 | 43 | 3 |
| 5 | 38.89 | 82 | 0 | | 22.22 | 21 | 0 |
| 6 | 61.11 | 98 | 6 | | 27.78 | 30 | 2 |
| 7 | 50.00 | 52 | 5 | | 33.33 | 25 | 4 |
| 8 | 38.89 | 50 | 3 | | 38.89 | 45 | 1 |
| 9 | 44.44 | 48 | 1 | | 38.89 | 42 | 2 |
| 10 | 50.00 | 51 | 3 | | 44.44 | 38 | 3 |

The traditional method of conducting inspection on a software artifact is to do it in teams of different sizes. However, it is not possible to gather the reviewers into teams online, since they did not meet face-to-face. Therefore, nominal teams' selection as is usually done in this area of research was used.

Nominal teams consist of individual reviewers or inspectors who do not communicate with each other during inspection work. Nominal teams can help to minimize inspection overhead and to lower inspection cost and duration [30]. The approach of creating virtual teams or nominal teams has been used in other studies as well, [30, 6]. An advantage of nominal inspections is that it is possible to generate and investigate the effect of different team sizes. A disadvantage is that no effects of the meeting and possible team synergy are present in the data [3]. The rationale for the investigation of nominal teams is to compare nominal inspections with the real world situation where teams would be formed without any re-sampling. There are many ways by which nominal teams can be created. Aybüke, *et al*, [3] suggest creating all combinations of teams where each individual reviewer is included in multiple teams, but this introduces dependencies among the nominal teams. They also suggest randomly create teams out of the reviewers without repetition and using bootstrapping as suggested by Efron and Tibshirani [11], in which samples are statistical drawn from a sample space, followed by immediately returning the sample so that it possibly can be drawn again. However, bootstrapping on an individual level will increase the overlap if the same reviewer is chosen more than once in a team.

In this study, four different nominal teams were created each for the Ad hoc and Checklist reviewers. The first reviewers form the team of size 1, next two form the teams of size two. So we have Teams of 1-person, 2-person, 3-person and 4-person. Table 2 gives the mean aggregate values of results obtained with the nominal teams.

**Table 2: Mean Aggregate Results for Nominal Team Sizes**

| **Ad hoc Inspection** | | | | | **Checklist Inspection** | | |
|---|---|---|---|---|---|---|---|
| **Nominal Team size** | **DE (%)** | **T (mins)** | **FP** | | **DE (%)** | **T (mins)** | **FP** |
| 1 | 27.78 | 31.0 | 1.0 | | 44.44 | 80.00 | 5.00 |
| 2 | 47.22 | 64.5 | 4.5 | | 30.56 | 47.50 | 4.50 |
| 3 | 50.00 | 76.7 | 2.7 | | 24.06 | 31.33 | 1.67 |
| 4 | 46.11 | 50.3 | 3.0 | | 38.89 | 37.50 | 2.50 |

Table 2 shows that 42.8% and 34.5% of the defects were detected by the ad hoc and checklist reviewers respectively in the inspection experiment.

The aggregate mean effort taken in minutes by the ad hoc reviewers was 55.62 ± 9.82 SEM minutes while the checklist reviewers take 49.08 ± 1 0.83 SEM minutes. SEM means Standard Error of Means. The aggregate mean false positives reported by the reviewers were 2.80 ± 0.72 SEM and 3.42 ± 0.79 SEM respectively for the ad hoc and checklist reviewers.

Fig. 2 shows the chart of defect detection effectiveness of the different teams in each of the defect detection method groups.

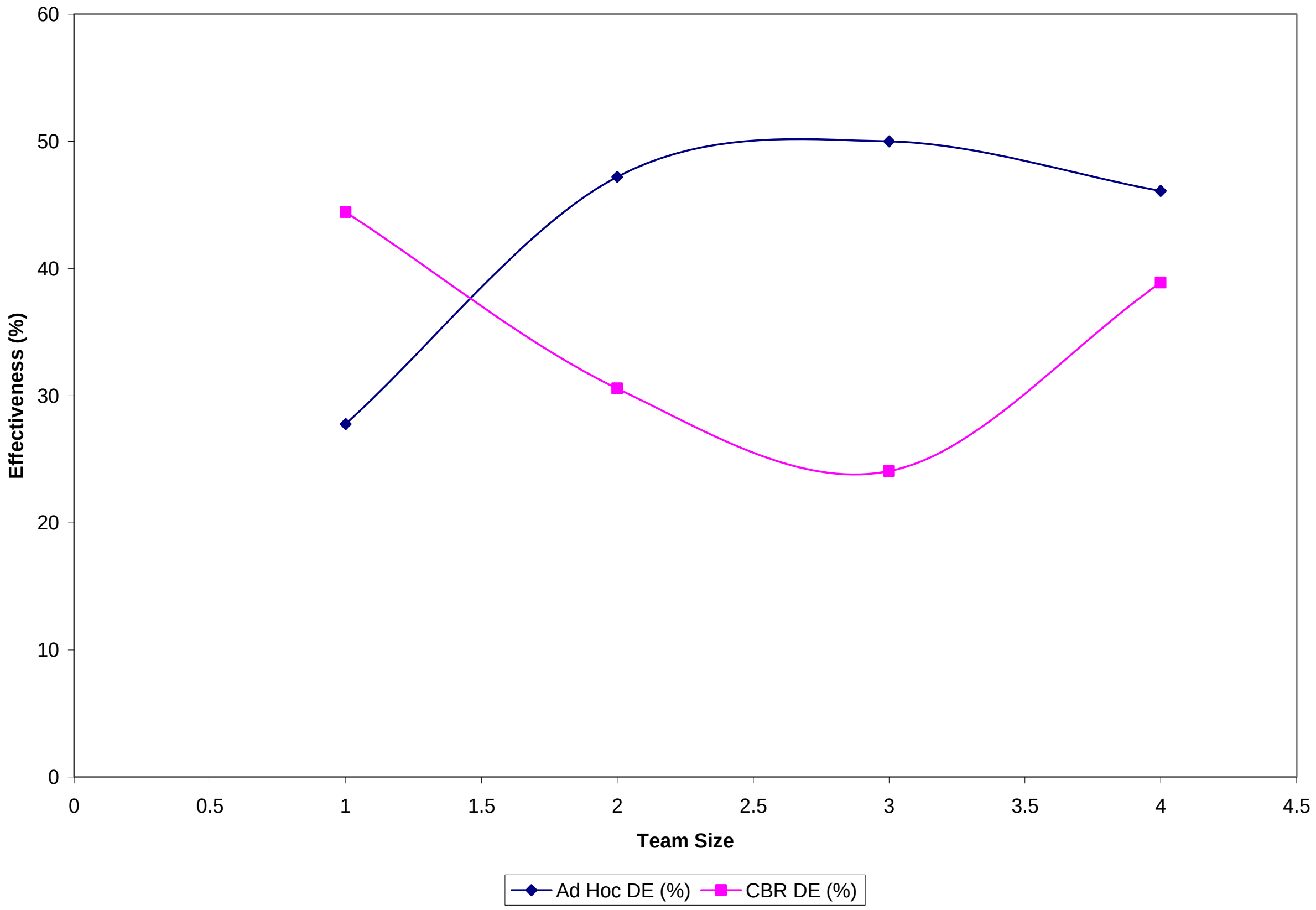

**Fig. 2: Chart of Defect Detection Effectiveness of Reviewers Against The Nominal Team Sizes**

Fig. 2 shows that the effectiveness of Ad hoc reviewers rises steadily and positively with team size, with peak value recorded on team size 3. However, checklist reviewers take a negative trend in that their effectiveness decreases with team size up to team size 3 before rising again on team 4.

Effort in terms of time taken by reviewers in minutes also takes the same shape as shown in Fig. 3.

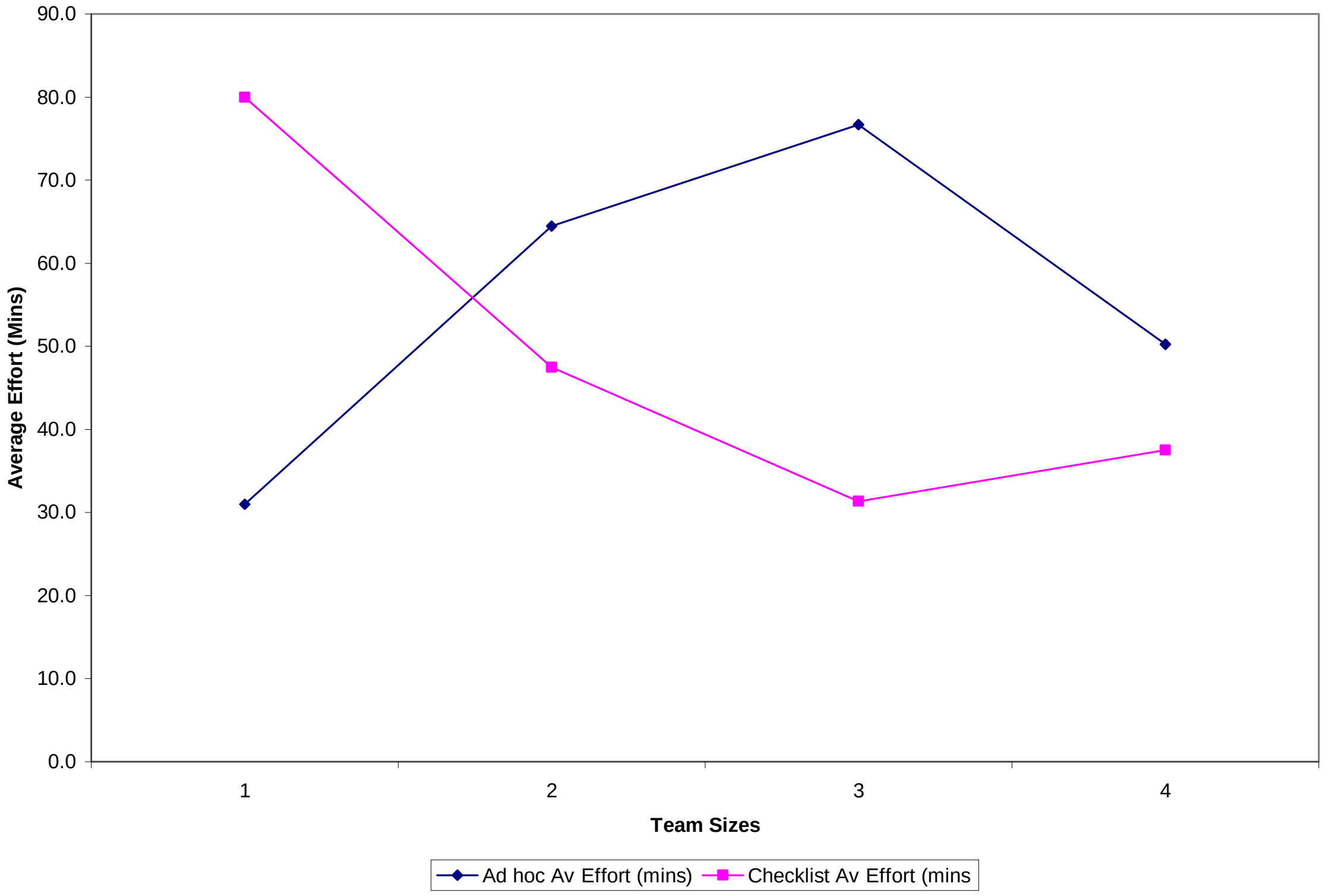

**Fig. 3: Chart of Effort (Mins) against the Nominal Team Sizes**

Fig. 4 shows the mean aggregate false positives reported by the reviewers in the experiment.

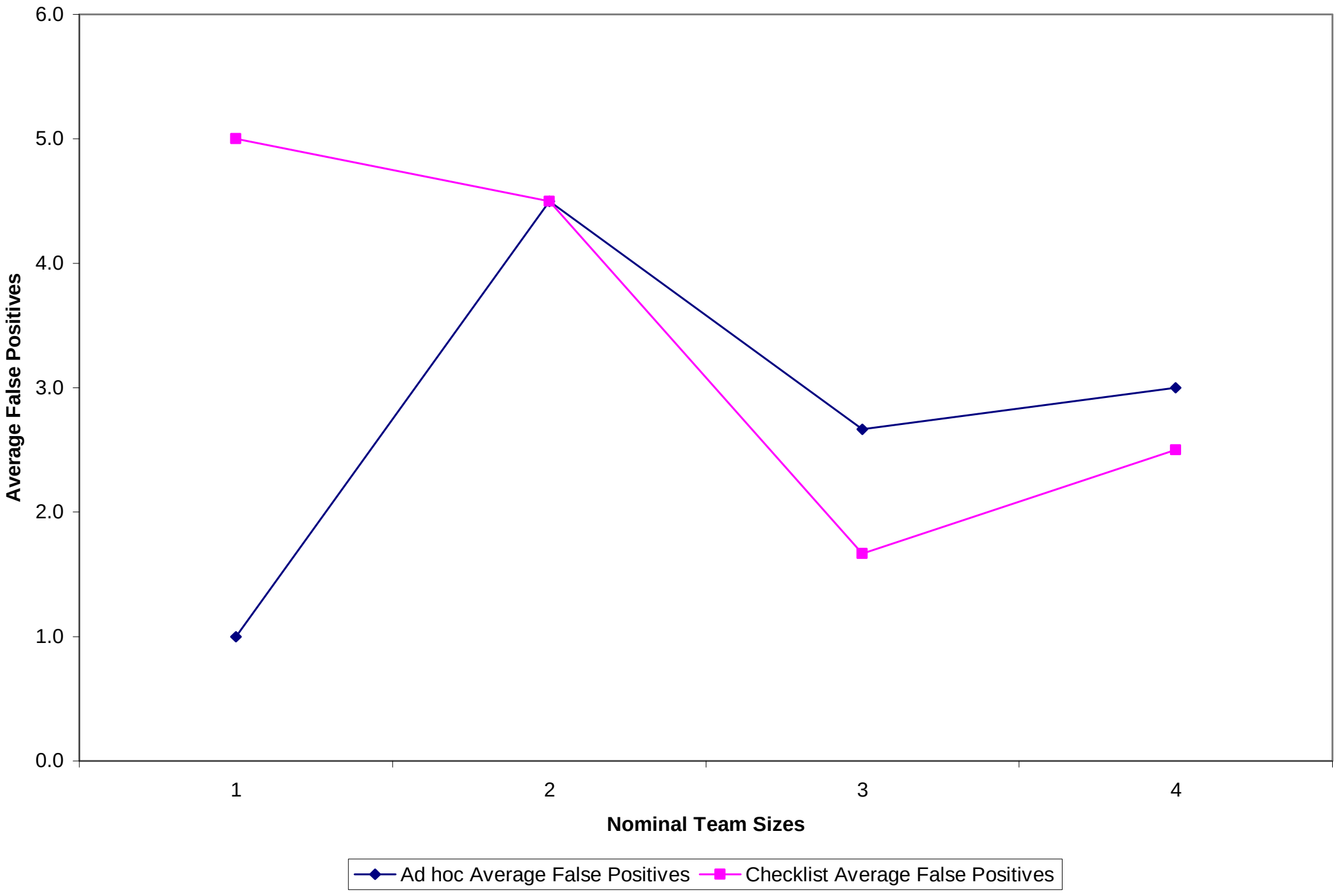

**Fig. 4: Average False Positives against the Nominal Team Sizes**

Examining the false positive curves in Fig. 4 critically, we could see that the shapes follow more or less trend with what are obtained on defect detection effectiveness and effort.

**Further Statistical Analyses**

Table 3 shows the results of major statistical tests performed on the data obtained in this experiment. Independent T-test was used for the analyses since different subjects were involved in the experiments and the experiments were carried out independently.

**Table 3: Major Statistical Test Results**

| Hypothesis Tested | p-value | Correlation | Decision |
|---|---|---|---|
| $H_o$: There is no significant difference between the effectiveness of reviewers using Ad hoc and Checklist techniques in distributed *code inspection.* | 0.267 | - 0.83 | $H_o$ accepted |
| $H_o$: There is no significant difference between the effort taken by reviewers using Ad hoc and Checklist techniques in distributed *code inspection.* | 0.670 | - 0.85 | $H_o$ accepted |
| $H_o$: There is no significant difference between the false positives reported by reviewers using Ad hoc and Checklist techniques in distributed *code inspection.* | 0.585 | - 0.15 | $H_o$ accepted |

Even though the shapes of the curves obtained on the experiment indicates differences in the defect detection effectiveness, the effort taken and the false positives reported by the reviewers, statistical tests conducted show that there are no significant differences in the defect detection effectiveness (p = 0.267), the effort taken (p = 0.670) as well as false positives reported by the reviewers (p = 0.585) for the two defect detection techniques understudied. Null hypotheses are thus accepted for all the tests.

Correlation coefficients are highly strong for defect detection effectiveness and effort / time taken albeit in negative directions as depicted in the charts in Figs. 2 and 3. There is a very weak negative correlation coefficient in the false positives reported by the reviewers.

## 5. Discussion of Results

Results from the groupware experiment show that there is no significant difference between the Ad Hoc and Checklist based reviewers in terms of the parameters measured – defect detection effectiveness, effort and false positives reported. Aggregate mean values of defect detection effectiveness and effort are slightly higher for Ad hoc reviewers while aggregate mean false positives result is slightly higher for Checklist based reviewers. About 43 and 35% of the defects were detected by the reviewers using ad hoc and checklist reading techniques respectively.

Our results are in consonance with some related works in the literatures. To mention a few, Porter and Votta [1] on their experiment for comparing defect detection methods for software requirements inspections show that checklist reviewers were no more effective than Ad hoc reviewers, and that another method, the scenario method, had a higher fault detection rate than either Ad hoc or Checklist methods. However, their results were obtained from a manual (paper-based) inspection environment.

Lanubile and Visaggio [21] on their work on evaluating defect detection techniques for software requirements inspections, also show that no difference was found between inspection teams applying Ad hoc or Checklist reading with respect to the percentage of discovered defects. Again, they conducted their experiment in a paper-based environment.

Nagappan, *et al*, [26] on their work on preliminary results on using static analysis tools for software inspection made reference to the fact that inspections can detect as little as 20% to as much as 93% of the total number of defects in an artifact. Briand, *et al*, [5] reports that on the average, software inspections find 57% of the defects in code and design documents.

In terms of percentage defects detected, low results were obtained from the experiment compare to what obtains in some related works. For instance, Giedre *et al*, [12] results from their experiment to compare checklist based reading and perspective-based reading for UML design documents inspection shows that Checklist-based reading (CBR) uncovers 70% in defect detection while Perspective – based (PBR) uncovers 69% and that checklist takes more time (effort) than PBR.

The implication of these results is that any of the defect detection reading techniques, Ad hoc or Checklist, could be conveniently employed in software inspection depending on choice either in a manual (paper-based) or tool-based environment; since they have roughly the same level of performance.

## 6. Conclusion and Recommendations

In this work we demonstrate the equality of ad hoc and checklist based reading techniques that are traditionally and primarily used as defect reading techniques in software code inspection; in terms of their defect detection effectiveness, effort taken and false positives. Our results show that none of the two reading techniques outperforms each other in the tool-based environment studied.

However, results in this study need further experimental clarifications especially in industrial setting with professionals and large real-life codes.

ix 1

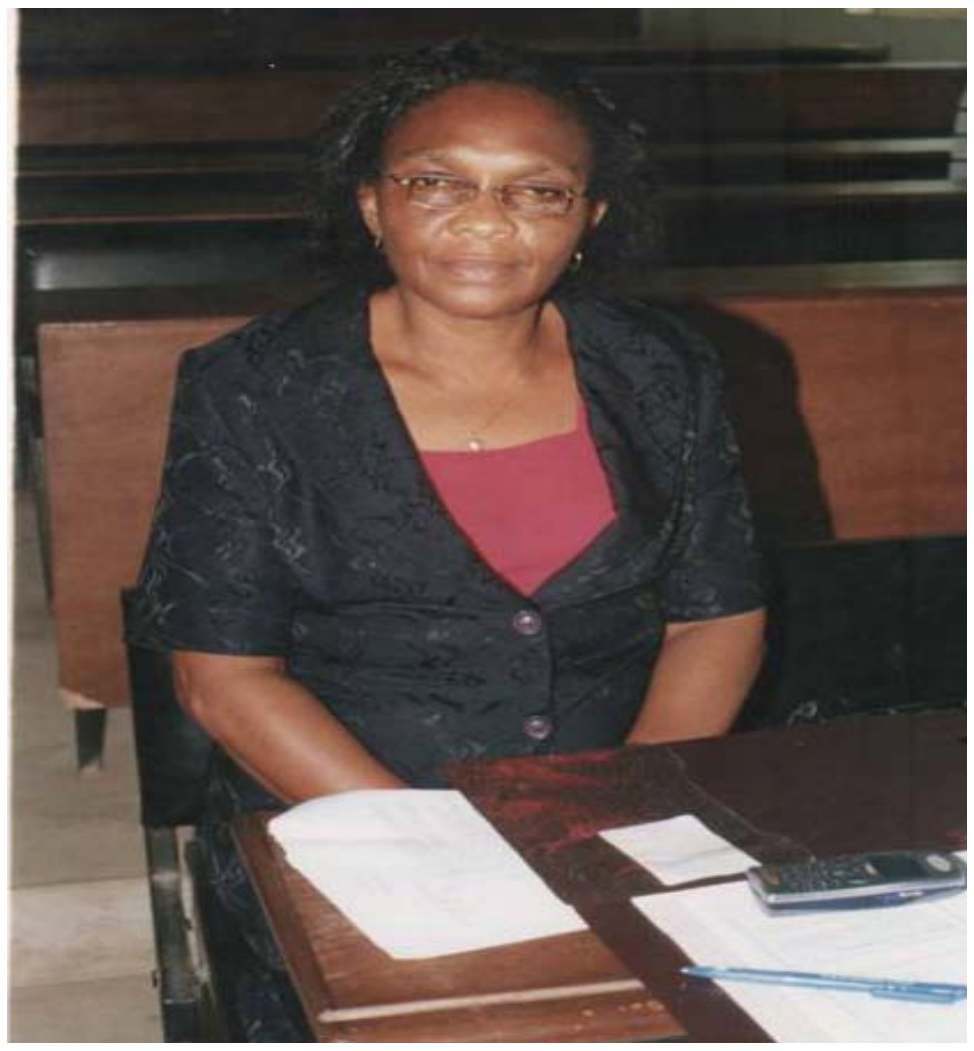

Adenike OSOFISAN is currently a Reader in the Department of Computer Science, University of Ibadan, Nigeria. She had her Masters degree in Computer Science from the Georgia Tech, USA and PhD from Obafemi Awolowo University, Ile-Ife, Nigeria. Her areas of specialty include data mining and communication.

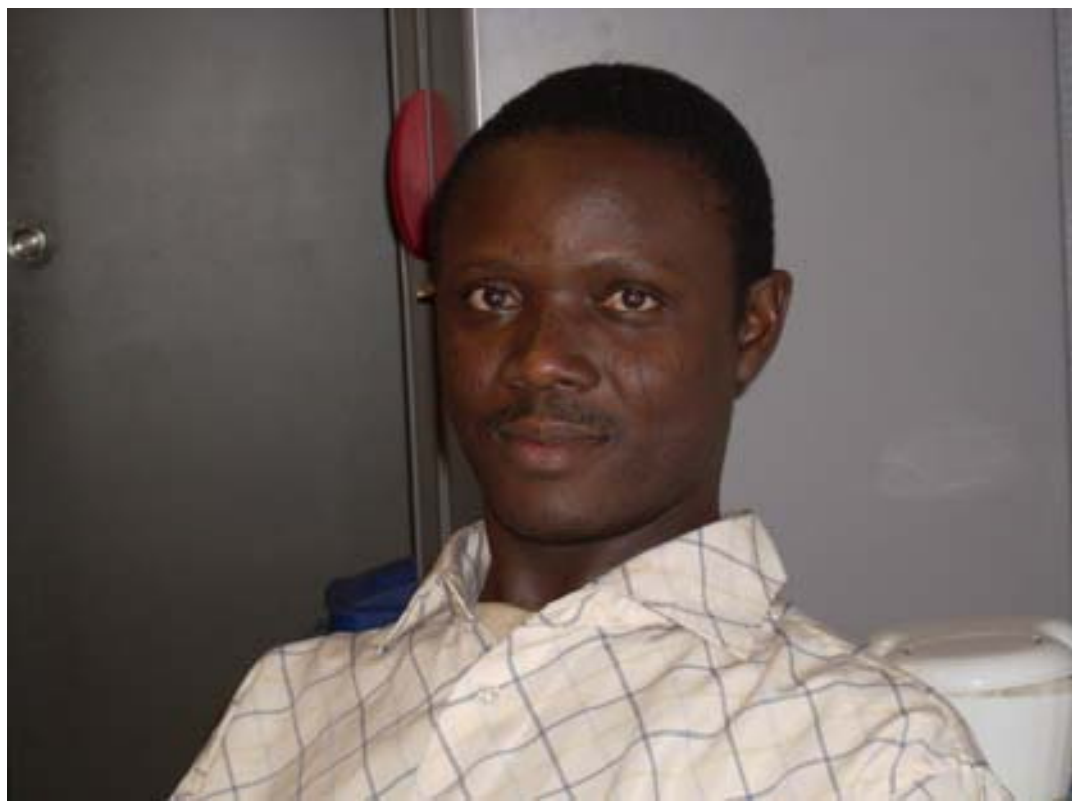

Solomon Olalekan AKINOLA is currently a PhD student of Computer Science in the University of Ibadan, Nigeria. He had his Bachelor of Science in Computer Science and Masters of Information Science in the same University. He specializes in Software Engineering with special focus on software inspection.